\documentclass[preprint,NumberedRefs]{JASA}

\usepackage{algpseudocode}
\usepackage{color}
\usepackage{mathtools}
\usepackage{lineno,hyperref}

\begin{document}

\nolinenumbers

\title{Acoustic radiation torque on a particle in a fluid: an angular spectrum based compact expression.}

\author{Zhixiong Gong}
\email{Corresponding author: zhixiong.gong@iemn.fr}
\author{Michael Baudoin}
\altaffiliation{Institut Universitaire de France, 1 rue Descartes, 75005 Paris, Email: michael.baudoin@univ-lille.fr}
\affiliation{Univ. Lille, CNRS, Centrale Lille, Yncréa ISEN, Univ. Polytechnique Hauts-de-France, UMR 8520 - IEMN, F- 59000 Lille, France}

\preprint{Gong \& Baudoin, JASA}		

\date{\today} 

\begin{abstract}

\nolinenumbers

In this work, we derive a set of compact analytical formulas expressing the three-dimensional acoustic radiation torque (ART) exerted on a particle of arbitrary shape embedded in a fluid and insonified by an arbitrary acoustic field. This formulation enables direct computation of the ART from the angular spectrum based beam shape coefficients introduced by Sapozhnikov \& Bailey [J. Acoust. Soc. Am. 133, 661–676 (2013)] and the partial wave coefficients. It is particularly well suited to determine the ART exerted on a particle when the acoustic field is known in a source plane.

\end{abstract}

\maketitle
\nolinenumbers
\section{\label{sec:Introduction}Introduction}
The acoustic radiation torque (ART) exerted by an arbitrary acoustic field on a particle can in general be decomposed into three contributions \cite{baresch2018orbital}: one resulting from the incident wave scattering by the particle, one induced by absorption of the acoustic field by the particle \cite{gong2019reversals} and one resulting from the wave absorption in the viscous boundary layer surrounding the particle \cite{prl_wang_1977,jasa_busse_1981,zhang2014acousticX1}. 
All these contributions are nonlinear second order effects, and not first order as suggested by the misleading title in Ref. \citep{prl_wang_1977}.
In addition, the particle can also be set in rotation by the so-called Eckart streaming \cite{pr_eckart_1948} - a flow resulting from the thermo-viscous absorption of the wave in the bulk of the fluid - in particular when the incident beam is carrying angular momentum \cite{prl_anhauser_2012,pre_riaud_2014,baresch2018orbital}. The ART can be calculated by transferring the integration of time-averaged stress tensor of angular momentum flux over the particle surface to a far-field spherical surface centered in the mass center of the particle, as first demonstrated by Maidanik \cite{maidanik1958torques,zhang2011acousticX2}.

Based on this idea, Zhang $\&$ Marston \cite{zhang2011angular} derived a compact formula of the axial ART ($T_z$) acting on an axisymmetric object centered on the axis of a cylindrical acoustical vortex beam beyond the paraxial approximation. The significance of energy dissipation in the vortex beam case was originally discussed in the paraxial limit \cite{hefner1999acousticalA1}. 
Zhang \& Marston showed that, in this configuration, the scattering contribution vanishes and the ART is proportional to the absorbed power ($P_{abs}$) with a factor $ M/ \omega$, with $M$ the beam’s topological charge and $\omega$ the angular frequency, \textit{i.e.}, $T_z = P_{abs} M/\omega$.  For a sphere in a Bessel vortex beam, the relation between dissipation and scattering was analyzed by Zhang \& Marston with an explicit expression of $P_{abs}$ \cite{zhang2011geometricalA2,zhang2011acousticA3}.
The theory applies for an elastic sphere in an inviscid fluid, and is also applicable for a sphere embedded in a weakly viscous fluid by modifying the scattering coefficients of the sphere \cite{zhang2011angular,baresch2018orbital,zhang2014acousticX1,zhang2018reversalsX3}. However, this theory is limited \cite{gong2019reversals} and cannot address the following situations: (i) non-axisymetric beams acting on a sphere (e.g. offset incidence of vortex beam on a sphere or oblique incidence on a spheroid and cylinder), (ii) non-axisymmetric objects with respect to the incident direction (e.g., broadside incidence on a spheroid), and (iii) multiple particles. 

In 2012, Silva \textit{et al.} \cite{silva2012radiation} used a spherical wave expansion of the incident and scattered beam -- the so called Multipole Expansion Method (MEM) -- to determine ART formulas applied on an arbitrary located sphere insonified by an arbitrary incident field in terms of the incident and scattered Beam Shape Coefficients (BSC, the coefficients corresponding to the projection of the wavefields on the spherical wave basis). These ART formulas were recovered later on by Gong \textit{et al} and extended to the case of arbitrary shape particles by using the T-matrix method. These theoretical developments were used to explore physical mechanisms at the origin of the three-dimensional (3D) torque reversal  \cite{gong2019reversals}.

But one major difficulty with the MEM is to calculate the BSC for an arbitrary field and an arbitrary located sphere. Baresch et al. \cite{baresch2013three} demonstrated that when the beam shape coefficients are known for a specific sphere location, they can be determined for any configuration by translating and rotating the spherical basis with some numerical toolbox. Also, for specific wavefields it is possible to determine analytical expressions of the BSC for an arbitrary located particle. The most simple configuration is the plane wave, since in this case the beam shape coefficients do not depend on the sphere location owing to the wavefield symmetry. The case of cylindrical Bessel beam was treated by Gong et al., who derived analytical expressions of the BSC for off-axis arbitrary incidence \cite{gong2017multipole} using Graf’s addition theorem. This expression was recovered later on by Zhang et al. \cite{zhang2018generalX0} using a different method. Nevertheless, all these methods are difficult to set into practice when analytical expressions of the incident beam are not known. Recently Zhao et al. \cite{Thomas2019computation} evaluated three methods to determine the Beam shape coefficients to compute the force for arbitrary (and in particular experimentally measured) acoustic fields: The first one relies on the orthogonality property of the spherical harmonics. From this property, the BSC can be obtained from a scalar product (integral over a spherical surface) of the wavefield  and the spherical harmonics. This approach was investigated by Silva to solve the off-axis scattering problem \cite{silva2011off}. Nevertheless, this method (i) requires to know or measure the value of the acoustic field over a sphere surrounding the insonified particle and (ii) was shown to induce fluctuation when calculating the acoustic radiation force and might lead to similar issued when computing the ART. The second method relies on the knowledge/measurement of the acoustic field at random points in a spherical volume and on the resolution of the inverse problem by a sparse approach. 
Finally the third method is based on the decomposition of the incident field into a sum of plane waves using the angular spectrum method  (ASM) introduced by Sapozhnikov \& Bailey  \cite{sapozhnikov2013radiation}  for the calculation of the acoustic radiation force. The major advantages of this last method are that (i) only the knowledge of the field in one plane is required (though the source can be plane or curved) and (ii) that the ASM-based BSC can be computed from simple integration of the angular spectrum (the 2D spatial Fourier Transform of the field in the reference plane) over a disk in the reciprocal space. Hence it is easy to set in practice, especially when using planar holographic transducers able to produce complex fields, such as acoustical vortices \cite{pre_jimenez_2016,prap_riaud_2017,apl_jimenez_2018,baudoin2019folding} or when the acoustic field can be measured in a reference plane \cite{melde2016holograms,Thomas2019computation}. 

In this paper, we derive compact analytical formula expressing  the torque applied by an arbitrary field on an arbitrary located particle of arbitrary shape and size as a function of the ASM-based beam shape coefficients. The formula are validated through comparison with previous results obtained  by Gong et al. \citep{gong2019reversals} with the MEM for an off-axis viscoelastic sphere insonified by a cylindrical Bessel beam. Note that here (see Appendix C), the BSC for cylindrical Bessel beam of off-axis arbitrary incidence are recovered with the ASM "method. Finally, the potential of this approach is illustrated by calculating the Torque applied on a $5 \, \mu$m particle insonified by a one-sided focused vortex beam produced by a plane active holographic transducer similar to the one used by Baudoin et al. to trap microparticles \cite{baudoin2019folding} and cells \cite{baudoin2020naturecell}.

\section{\label{sec: Theoretical method} Angular spectrum based ART formulas}
In this section, we give a brief overview of the main steps leading to the derivation of the angular spectrum based ART formulas (with the same notations as in Ref. \citep{gong2019reversals}). The acoustic radiation torque exerted by an acoustic field on a particle can be calculated by transferring the integration of the time-averaged ($\langle \cdot \rangle$) stress tensor of the angular momentum flux over the particle surface to a far-field spherical surface $S_0$ \cite{maidanik1958torques,silva2012radiation,gong2019reversals,zhang2011acousticX2}. 
Based on the divergence theorem, the integral expression of the ART is:
\begin{equation}
\mathbf{T}=-\rho_{0} \iint_{S_{0}}\langle L\rangle \mathbf{r} \times d \mathbf{S}-\rho_{0} \iint_{S_{0}}\langle(\mathbf{r} \times \mathbf{u}) \mathbf{u}\rangle d \mathbf{S},
\label{torque general}
\end{equation}
where $\langle L \rangle = \langle 1/2 \; \mathbf{u}\cdot \mathbf{u}) - p^2 / (2 \rho_0 c_0^2) \rangle$ is the time-averaged acoustic Lagrangian, $\textbf{r}$ is the field point,  $p$ is the total acoustic pressure field (incident + scattered), $\textbf{u}$ is the total acoustic velocity vector, $\rho_0$ is the fluid density at rest, $c_0$ is the fluid sound speed, and $d \mathbf{S}=\mathbf{n} \cdot r^{2} \sin \theta d \theta d \varphi$ is the differential surface in the far field with $\textbf{n}$ the outward unit normal vector.  
If $S_0$ is a sphere whose center coincide with the referential center, then $\mathbf{r} \times \mathbf{n} = r \mathbf{n} \times \mathbf{n} = \mathbf{0}$ and the first term in Eq. (\ref{torque general}) vanishes. 
Then, the incident (index "i") and scattered (index "s") acoustic velocity $\mathbf{u}$ and pressure $p$ fields can be described in terms of acoustical potentials $\Phi$ as:
\begin{equation}
\mathbf{u}_{i, s}=\nabla \Phi_{i, s} \mbox{ and } p_{i, s}=i \omega \rho_{0}  \Phi_{i, s},  \label{potential}
\end{equation}
with $i$ is the imaginary unit and $\omega$ the angular frequency. 
Hence, the ART expression in Eq. (\ref{torque general}) can be written in terms of the incident ($\Phi_{i}$) and scattered ($\Phi_{s}$) velocity potentials as
\begin{equation}
\mathbf{T}=\frac{\rho_{0}}{2} \operatorname{Im}\left\{\iint_{S_{0}}\left(\frac{\partial \Phi_{i}^{*}}{\partial r} \mathbf{L} \Phi_{s}+\frac{\partial \Phi_{s}^{*}}{\partial r} \mathbf{L} \Phi_{i}+\frac{\partial \Phi_{s}^{*}}{\partial r} \mathbf{L} \Phi_{s}\right) d S\right\}.
\label{torque in potential} 
\end{equation}
where ``Im'' designates the imaginary part, the star superscript the complex conjugate and $\mathbf{L}=-i(\mathbf{r} \times \nabla)$ is the angular momentum operator, with its components in the three directions and their recursion relations with normalized spherical harmonics given in detail in Appendix.

Now, assuming that the incident pressure field is known in a plane defined as $z = 0$, $p_i |_{z=0} = p_i(x,y,0)$, the angular spectrum  $S(k_x,k_y)$ of the acoustic field (that is nothing but the 2D spatial Fourier transform of the complex temporal harmonic amplitude of the field in this plane) reads:
\begin{equation}
\begin{aligned}
S\left(k_{x}, k_{y}\right)=\int_{-\infty}^{+\infty} \int_{-\infty}^{+\infty}  p_{i}(x, y, 0) e^{-i k_{x} x - i k_{y} y} d x d y
\end{aligned},
\label{S(kx,ky)}
\end{equation}
with $x$ and $y$ the cartesian coordinates in the plane, $k_x$ and $k_y$ the wavenumber components in $x$ and $y$ directions. Then, the field at any point can be calculated by propagating each plane wave composing the source plane up to the target point $(x,y,z)$:
\begin{equation}
\begin{multlined}
p_i(x,y,z) = \frac{1}{4 \pi^2} \\ \times \iint_{k_x^2 + k_y^2 \leq k^2} S(k_x,k_y) e^{i k_{x} x + i k_{y} y + i \sqrt{k^2  - k_x^2 - k_y^2} z}d k_{x} d k_{y}.
\end{multlined}
\end{equation}
where $k = \omega / c_0$ is the wave number in fluid. In this way, the acoustic field is decomposed into an infinite sum of plane waves and the angular spectrum  $S(k_x,k_y)$ characterizes the relative magnitude of each plane wave. The next step is to solve the scattering problem. For this purpose, this plane wave decomposition must be turned into a spherical wave decomposition, more suitable to solve the scattering problem (see Sapozhnikov \& Bailey \cite{sapozhnikov2013radiation}):
\begin{equation}
p_{i}  =  \frac{1}{ \pi } \sum_{n=0}^{\infty} \sum_{m=-n}^{n} i^{n}  H_{n m} j_{n}(k r)  Y_{n m}(\theta, \varphi) , 
\label{incident_pressure}
\end{equation}
where $Y_{n m}(\theta, \varphi)$ are the spherical harmonics and the $H_{nm}$ represent the respective weight of each spherical wave and hence are nothing but ASM-based beam shape coefficients:
\begin{equation}
\begin{aligned}
H_{n m}=\iint_{k_{x}^{2}+k_{y}^{2} \leq k^{2}}  S\left(k_{x}, k_{y}\right) Y_{n m}^{*}\left(\theta_{k}, \varphi_{k}\right)d k_{x} d k_{y}
\end{aligned},
\label{Hnm}
\end{equation}
with $\cos\theta_k = [1-\left(k_{x}^{2}+k_{y}^{2}\right) / k^{2}]^{1/2}$ and $\varphi_k = \arctan \left(k_{y} / k_{x}\right)$. This expression results from the known decomposition of a plane wave into a sum of spherical waves. 
Analytical solutions of the scattering problem for spheres embedded in a fluid are known in many cases including rigid\cite{marston2007scattering}, elastic \cite{faran1951sound} or visco-elastic particles \cite{baresch2018orbital,gong2019reversals}. For non-spherical particles, the scattering problem can be handled with the so-called T-matrix method \cite{waterman1969new,bostrom1984scattering,varadan1988comments,lim2015more,gong2016arbitrary,gong2017analysis}.
Assuming prior knowledge of the scattering coefficients, the scattered field can be written under the form:
\begin{equation}
p_{s} =  \frac{1}{\pi } \sum_{n=0}^{\infty} \sum_{m=-n}^{n} i^{n}  H_{n m} A_{nm} h_{n}^{(1)}(k r) Y_{n m}(\theta, \varphi) , \label{scattered_pressure}
\end{equation}
with $A_{nm}$ the partial wave coefficients which only depend on the index $n$ for a spherical shape, having $A_n =(s_n-1)/2$ with $s_n$ the scattering coefficients, and depend on both $n$ and $m$ for non-spherical shapes \cite{gong2019reversals}.

Now the incident and scattered pressure fields are given in terms of the BSC $H_{nm}$ based on the angular spectrum method \cite{sapozhnikov2013radiation}. 
The incident ($\Phi_{i}$) and scattered ($\Phi_{s}$) velocity potentials can be easily obtained by using the second equation of Eq. (\ref{potential}), which can then be substituted into Eq. (\ref{torque in potential}). 
Since the integral is performed on the far field surface $S_0$, the asymptotic expressions of Bessel functions can be used [ Eqs. (\ref{asymptotic expressions}) in appendix], which combined to the recursion relation of Bessel functions [Eqs.(\ref{jn recursion}) in Appendix] leads to the following expression of the ART expression in terms of $H_{nm}$ and $Y_{n}^{m}$

\begin{equation}
\begin{aligned}
\mathbf{T}=& -\frac{1}{2 \pi^{2} \rho_{0} k^{3} c_{0}^{2}} \operatorname{Re} \left\{ \sum_{n=0}^{\infty} \sum_{m=-n}^{n} \sum_{n=0}^{\infty} \sum_{m'=-n'}^{n'}\left(1+A_{n}^{m*}\right) A_{n'}^{m'}  \right.
\\
& \left. \times H_{n m}^{*}  H_{n' m'} \iint_{S_{0}}\left(Y_{n}^{m*} \mathbf{L} Y_{n'}^{m'}\right) \sin \theta d \theta d \varphi\right\},
\label{a_nm & H_nm}
\end{aligned}
\end{equation}

The final compact expression of the three-dimensional ART in terms of the $H_{nm}$ coefficients can be derived by using the recursion and orthogonality relations of the normalized spherical harmonics (see details in Appendix \ref{Appendix B}):

\begin{align}
& T_x=-\frac{1}{4 \pi^{2} \rho_{0} k^{3} c_0^{2}} \operatorname{Re}\left\{\sum_{n=0}^{\infty} \sum_{m=-n+1}^{n}
b_{n}^{m} C_{n}^{m} H_{n m}^{*} H_{n, m-1}\right\}, \label{Tx}
\\
& T_y=-\frac{1}{4 \pi^{2} \rho_{0} k^{3} c_0^{2}} \operatorname{Im} \left\{\sum_{n=0}^{\infty} \sum_{m=-n+1}^{n} b_{n}^{m} C_{n}^{m} H_{n m}^{*} H_{n, m-1}\right\}, \label{Ty}
\\
& T_z=-\frac{1}{2 \pi^{2} \rho_{0} k^{3} c_0^{2} } \operatorname{Re} \left\{\sum_{n=0}^{\infty} \sum_{m=-n}^{n} m D_{n}^{m} H_{n m}^{*} H_{n m}\right\}, \label{Tz}
\end{align}
with $b_{n}^{m} = \sqrt{(n-m+1)(n+m)}$, $C_{n}^{m} = A_{n}^{m-1}+2 A_{n}^{m-1} A_{n}^{m*} + A_{n}^{m*}$, $D_{n}^{m} = A_{n}^{m} + A_{n}^{m} A_{n}^{m*}$.
Note that the prior ART formulas by Silva \textit{et al} \cite{silva2012radiation} were not written in a compact form since they do not express the scattered BSC in terms of the product of the incident BSC and $A_n$ for a spherical shape and have index issues \cite{baudoin2020acoustic}. A thorough comparison between the present formula and the one obtained by Silva \textit{et al} \cite{silva2012radiation} is provided in Ref. \cite{gong2020equivalence}.

\section{\label{sec:validation} Validation of the angular spectrum based ART formulas for an off-axis visco-elastic sphere insonified by a Bessel beam.}
\begin{figure*}
\includegraphics[width=0.8\linewidth]{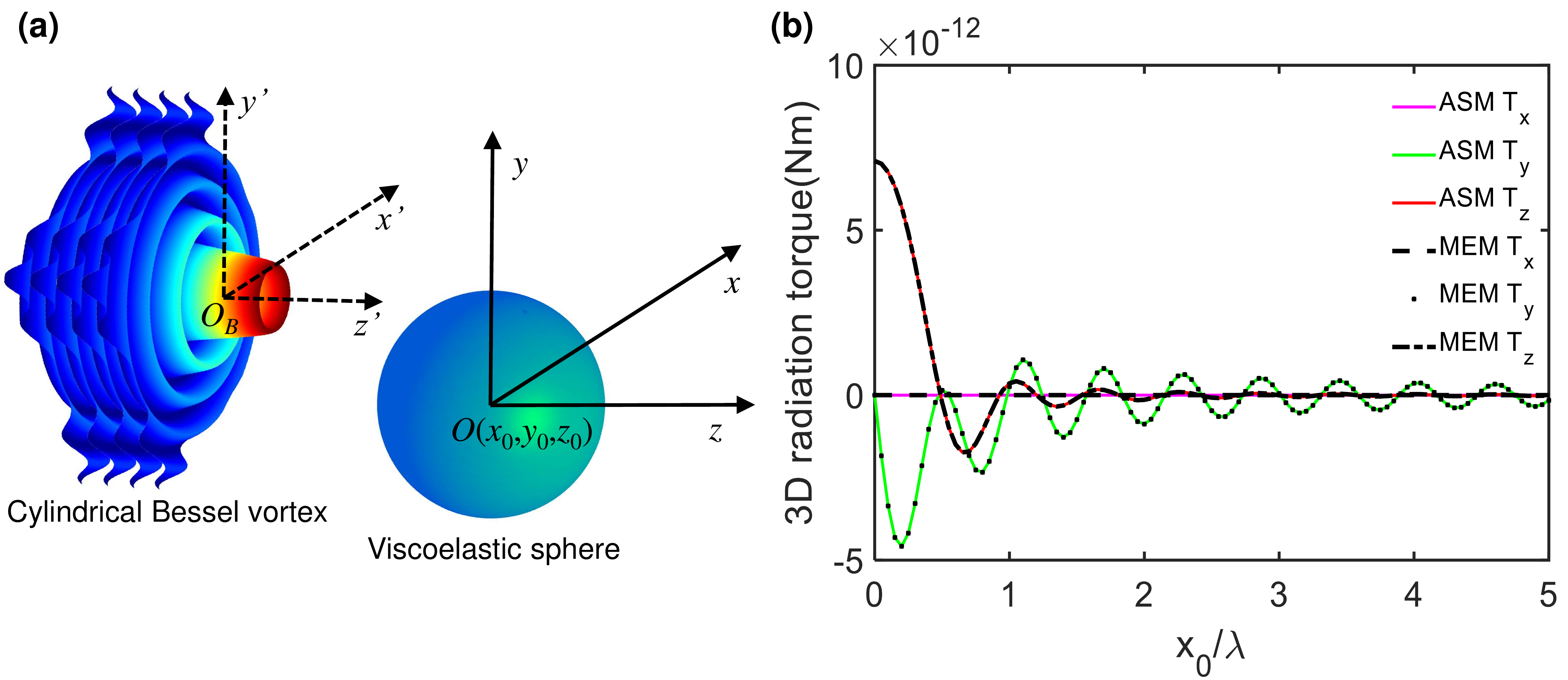}
\caption{\label{Fig1:three-dimensional ART} (color online) The 3 projections of the acoustic radiation torque ($T_x$, $T_y$, $T_z$) exerted on an off-axis viscoelastic sphere by a cylindrical Bessel vortex on a viscoelastic PE solid sphere is calculated by the present Angular Spectrum Method (ASM) (equations (\ref{Tx}-\ref{Tz})) and compared to results obtained with the Multipole Expansion Method (MEM) (Gong et al. \cite{gong2019reversals}). (a) Scheme of the simulated configuration. The topological order of the cylindrical Bessel beam is $M=1$ and the cone angle $\beta = 60^{\circ}$ (see ref. \cite{baudoin2020acoustic} for more details about acoustical vortices). The particle is moved away from the beam center along $x$ direction of a distance: $x_0 \in [0,5\lambda]$, with $\lambda$ the wavelength in the fluid. There is no offset along the $y$ direction ($y_0 = 0$) and since cylindrical Bessel are invariant along $z$, the position along this axis does not matter. (b) Figure comparing the values of the 3 projections of the acoustic radiation torque obtained with ASM and MEM, as a function of the particle dimensionless offset $x_{0} / \lambda$.}

\end{figure*}
To validate the ART expressions obtained with the angular spectrum method (ASM) in the previous section, the torque exerted on an off-axis viscoelastic sphere insonified in an inviscid fluid by a cylindrical Bessel vortex is calculated with Eqs. (\ref{Tx} -- \ref{Tz}) and compared with the results obtained with the  multipole expansion method (MEM) \cite{gong2017multipole} by Gong \textit{et al.}\cite{gong2019reversals}. 
Note that for a sphere in a vortex beam, there are two forms of rotations when the sphere is off the beam axis: (i) the orbital rotation of the sphere around the beam axis induced by the azimuthal force, and (ii) the spin rotation of the sphere around its own mass center by the torque. For a non-absorbing sphere located on the axis of the vortex, there is no axial torque following the theory proposed by Zhang \& Marston \cite{zhang2011angular} ($T_z = P_{abs} M/\omega$) since $P_{abs}=0$. 
In this configuration, analytical expression of the $H_{nm}$ coefficients is given by (see detailed analytical derivation with ASM in Appendix \ref{Appendix C2}):
\begin{equation}
\begin{multlined}
H_{n m}=4 \pi^2 \omega \rho_{0} \Phi_{0} \xi_{n m} \\
\times i^{M-m+1} P_{n}^{m}(\cos \beta) J_{m-M}\left(k_\perp R_0 \right)  e^{-i k_{z} z_{0}} e^{i(M-m) \varphi_{0}}
\label{H_nm CBB}
\end{multlined}
\end{equation}
where $\xi_{n m}=[(2 n+1)(n-m) !]^{1 / 2}[4 \pi(n+m) !]^{-1 / 2}$, $M$ is the topological charge of the Bessel beam, $\beta$ is the cone angle, $k_{\perp}=k \sin \beta$, $R_0 = \sqrt{x_{0}^{2}+y_{0}^{2}}$, and $x_0$, $y_0$ and $z_0$ are the offset along the $x$, $y$ and $z$ directions, respectively. 
The axial component of the wave number is $k_{z}=k \cos \beta$, and the original azimuthal angle is $\varphi_0 = \tan ^{-1}\left(y_{0} / x_{0}\right)$. 
Note also that Eq. (\ref{H_nm CBB}) obtained here with ASM is equivalent to previous theoretical expression obtained by Gong et al. \cite{gong2017multipole} and Zhang \cite{zhang2018generalX0}. By inserting Eq. (\ref{H_nm CBB}) into (\ref{Tz}), the expression of axial ART is verified equivalent to Eq. (15a) of Ref. \cite{zhang2018reversalsX3}.
In the simulations represented on Fig. \ref{Fig1:three-dimensional ART}, the topological charge of the cylindrical Bessel beam is $M=1$ with $\beta = 60^{\circ}$, the incident frequency is 1 MHz with pressure amplitude 1 MPa, and the particle radius is $a=180$ $\mu$m. 
For brevity, the acoustic parameters of the viscoelastic polyethylene (PE) sphere immersed in water are same as those used in Ref. \cite{gong2019reversals}, as given in Table \ref{table1}.
The particle is moved off the beam axis [see the schematic in Fig. (\ref{Fig1:three-dimensional ART}a)] along only $x$ direction with $x_0 \in [0,5\lambda]$, $y_0=0$ and $z_0 = 0$ where $\lambda=1.5$ mm is the wavelength in water. 
As observed in Fig. (\ref{Fig1:three-dimensional ART}b), the calculated three-dimensional ART with the ASM [see Eqs. (\ref{Tx}-\ref{Tz})] agree exactly with those by the MEM \cite{gong2019reversals,gong2020equivalence}. No computational error is expected between the two methods since the two theoretical expressions can be shown to be equivalent \cite{gong2020equivalence}. 
Note that the partial wave (or scattering) coefficients $A_{nm}$ for a viscoelastic sphere of both results are obtained based on the Kelvin-Voigt linear viscoelastic model \cite{gaunaurd1978theory} with the explicit expressions given in the Appendix of Ref. \cite{gong2019reversals}. Further confirmation of these formula is under way by comparing directly the analytical expressions of the two formulas \cite{gong2020equivalence}, i.e., the relation between the BSC can turn the ART formulas in Eqs. (\ref{Tx}-\ref{Tz}) into those by Silva \textit{et al} \cite{silva2012radiation} if index issues are improved.
Note that since the particle is moved off axis along the $x$ direction, the lateral ART $T_x$ always vanishes because of the symmetry. For the normal incidence ($x_0 = 0$ and $y_0$ = 0), only the axial ART due to acoustic absorption exists \cite{zhang2011angular,baresch2018orbital}.

\begin{table}[!htbp]
\small
  \caption{Acoustic parameters of particle materials and water. The absorption values for the polyethylene (PE) were derived from ultrasonic measurements made by Hartmann and Jarsynski\cite{hartmann1972ultrasonic} with the longitudinal and shear absorption per wavelength 0.4 dB and 1.2 dB, respectively, while for the polystyrene (PS) by Takagi \textit{et. al.}\cite{takagi2007relaxation} with the longitudinal and shear absorption coefficients 23 Np/m and 108 Np/m 
 at frequency $f=5$ MHz. Note that 1 Np = 8.6859 dB. The normalized longitudinal ($\gamma_p$) and shear ($\gamma_s$) absorption coefficients are normalized by the corresponding wave number ($k_{p,s}$) which are calculated by the longitudinal ($c_p$) and shear ($c_s$) velocities with $\omega=k_{p,s}c_{p,s}$.}
  \label{table1}
  \begin{tabular}{l c  c  c  c  c }
\hline
Material &  Density(kg/m$^3$) &  $c_p$(m/s) & $c_s$(m/s) & $\gamma_p$ & $\gamma_s$  \\
\hline 
PE & 957 & 2430 & 950 & 0.0074 & 0.022 \\

PS & 1050 & 2350 & 1100 & 0.0017 & 0.0038 \\

Water & 1000 & 1500 & ... & ... & ... \\
 \hline
  \end{tabular}
\end{table}

\begin{figure*}
\includegraphics[width=\linewidth]{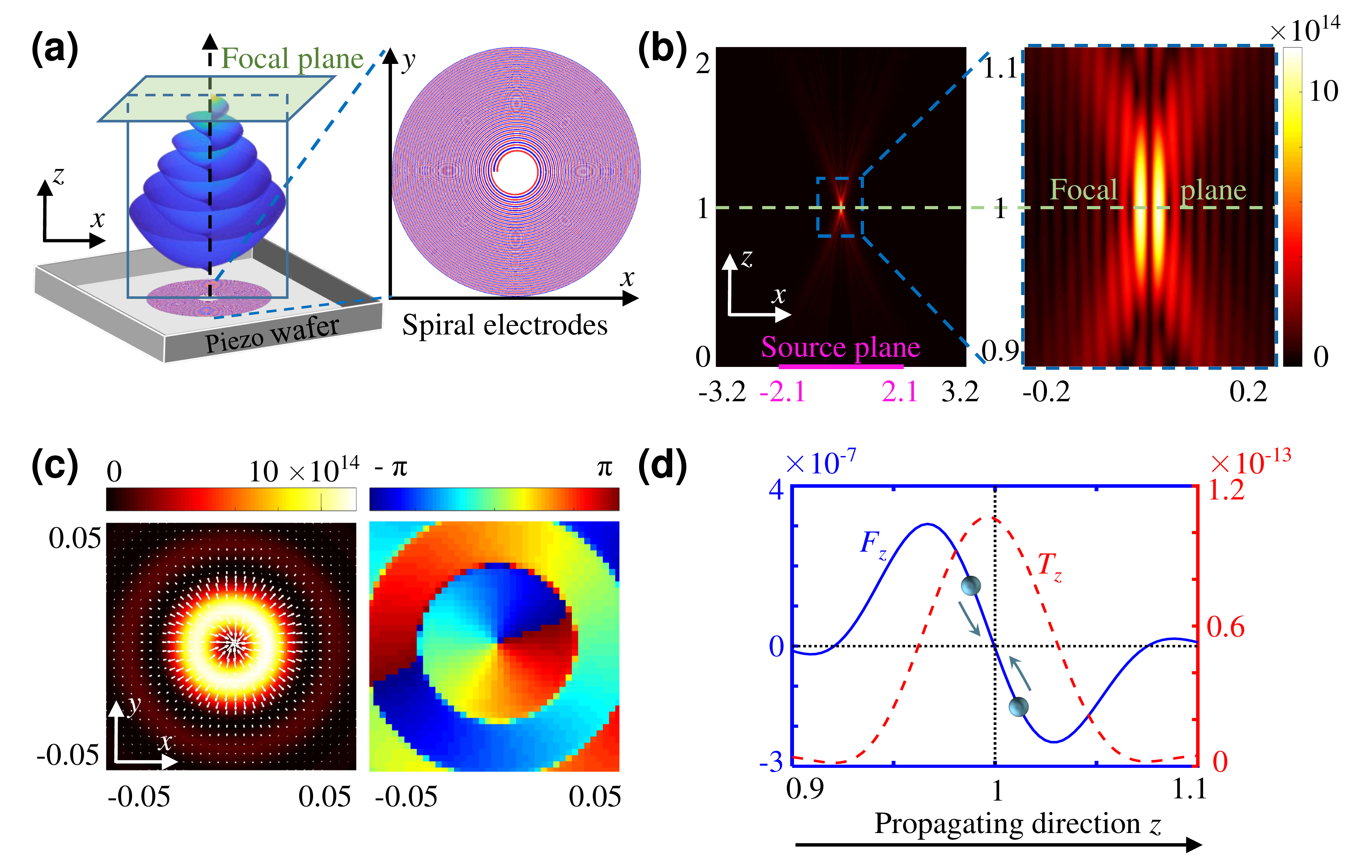}
\caption{\label{Fig3:ART in real field} (color online) (a) Schematic of a one-sided focused vortex synthesized (left) with a set of two spiraling electrodes of inverse polarity (right, red and blue). The focal plane is designed at $z=1$ mm with the source plane at $z=0$. The aperture of the spiral electrodes is $R_t=$ 2.1 mm. (b) Acoustic pressure magnitude square in the ($x,z$) plane (left) with a zoom near the focal point (right). The lengths along $x$ and $z$ are not scaled. (c) Acoustic pressure magnitude square and phase in the lateral focus plane $(x,y)$ at $z=1$ mm. The arrows on the first panel give the lateral force for a  viscoelastic PS particle with $a=5 \, \mu$m in radius in the focal plane, showing a slightly out-centered lateral trap. (d) The axial radiation force ($F_z$ in Newton) and torque ($T_z$ in N$\cdot$m) versus $z$ along the lateral trap axis.  The particle is trapped in 3D [see the first panel of (c) for lateral trap and the blue line for axial trap] and spins around its mass center located on the beam axis. The lengths are in the unit of mm in (b-d).}
\end{figure*}

\section{\label{sec:real field} Study of the ART exerted by a one-sided spherical vortex on a viscoelastic sphere}

To illustrate the potential of the present approach, we now apply our method to a case which cannot be treated analytically: the torque exerted on a $5 \, \mu$m viscoelastic polystyrene (PS) particle insonified by a $40$ MHz one-sided spherical vortex synthesized in an inviscid fluid by plane active spiraling transducers (Fig. \ref{Fig3:ART in real field}a) similar to the one used by Baudoin \textit{et al.} to trap microparticles \cite{baudoin2019folding} and cells \cite{baudoin2020naturecell}. The transducers are made of two active spiraling electrodes of inverse polarity whose equations are given in Ref. \cite{baudoin2019folding} exciting a piezoelectric wafer. 
In these simulations, the maximum radius of the transducer is $R_t=$ 2.1 mm and the transducer is designed to obtain a focal plane located at $z=1$ mm from the source plane, leading to a large aperture angle $65^{\circ}$ to obtain axial and hence 3D trapping capabilities. The acoustic pressure in the source plane $p_i(x,y,0)$ is approximated as having the same geometrical distribution as the active spiral electrodes as shown in Fig. \ref{Fig3:ART in real field}(a). That is to say, for the red electrode, the pressure has amplitude 1 MPa and phase 0, written as $p_i(x,y,0)=$ 1 MPa, while for the blue one, the pressure has amplitude 1 MPa and phase $\pi$, written as $p_i(x,y,0)=-$ 1 MPa. The rest of the domain has a  pressure equal to  $p_i(x,y,0)=$ 0. In the following computations, we take the physical domain as $3R_t \times 3R_t$ in $(x,y)$ plane and the fine mesh number 3000 with the physical space interval $R_t/1000$ to assure the convergence. Based on the two-dimensional fast Fourier transform, the angular spectrum is directly computed with the wave number interval $2\pi/(3R_t)$ and the range $2000\pi/R_t \times 2000\pi/R_t$ in $(k_x,k_y)$ plane.
The torque is computed with the formula provided in this paper while the force is computed using formulas by Sapozhnikov \& Bailey \cite{sapozhnikov2013radiation}. The field synthesized by these transducers computed with angular spectrum method is illustrated in Fig. \ref{Fig3:ART in real field}b for the ($x,z$) plane and Fig. \ref{Fig3:ART in real field}c for the ($x,y$) plane.  The lateral force applied in the focal plane on the particle is shown as arrows on Fig. \ref{Fig3:ART in real field}c (left). This figure shows that the particle trap position is a bit out-centered, which can be simply explained by the spiral finiteness. The axial force and torque is then calculated on the lateral trapping axis and represented in Fig. \ref{Fig3:ART in real field}d. This figure shows that spiraling transducers exhibit 3D trapping capabilities providing that the aperture is sufficient (here we chose an aperture of $65^{\circ}$). 
Note that the 3D trapping of 190 to 390 $\mu$m PS particles in the MHz range with a focused vortex was demonstrated first theoretically and then experimentally with a complex array of transducers in ref \cite{baresch2013spherical,baresch2016observation}. It also shows that the axial torque is maximum for a value slightly below the focal plane. 

\section{\label{sec:conclusion} Conclusions and discussions}

In summary, some compact angular spectrum based three-dimensional ART formulas are derived for a single particle immersed in an ideal fluid with no limitation to the particle size, particle shape and beam shape structure. 
The arbitrary acoustic field is taken as the superposition of plane waves, hence can be expanded based on the angular spectrum method \cite{sapozhnikov2013radiation}, which is quite practical for finite-aperture real sources. The ART on non-spherical shapes (e.g., spheroid and finite cylinder) can be calculated once the partial wave coefficients $A_{n}^{m}$ are obtained with proper methods, for example, the T-matrix method \cite{gong2019reversals,waterman1969new,bostrom1984scattering,varadan1988comments,lim2015more,gong2016arbitrary,gong2017analysis}.
The present theory are still practicable for an absorbing sphere in a viscous fluid if the absorption processes in the particle and viscous layer is accounted in the expression of scattering coefficients \cite{zhang2011angular,baresch2018orbital,zhang2014acousticX1,zhang2018reversalsX3}.  
In addition, the formulas can be used for multiple objects \cite{bostrom1980multiple,lopes2016acoustic,baresch2018orbital} which are all located inside the chosen far-field spherical shape so that the divergence theorem still holds for the derivation. The ART of particle in experimental sources can be evaluated by the measured acoustic field in the transverse plane as similar as the simulation of acoustic radiation force \cite{Thomas2019computation}.
By Combining with the three-dimensional acoustic radiation forces \cite{sapozhnikov2013radiation}, we can predict the dynamic motions of particles in real acoustic field with six degrees of freedom, i.e., three for translocations and three for spinning motions. It is noteworthy that for a particle located off the axis of a vortex beam, both three-dimensional radiation forces and torques are applied on the particle so that the particle could rotate around the beam axis (by the azimuthal component of radiation force) and its own center of mass (by the radiation torque).
Since particle in a vortex beam could be ejected out of the trap \cite{marzo2018acoustic}, this work can be used for theoretical guidance on parameters selection of acoustic sources for experimental designs, which can slow down the spinning motions by decreasing the ART, and meanwhile, keep the trapping by the acoustic radiation force. 
In addition, this work has the potential to dynamically control the rotation and translation of particles in manipulation devices in and beyond the long-wavelength regime \cite{inoue2019acoustical,gong2018Reversals}.
\begin{acknowledgments}
We acknowledge the support of the programs ERC Generator and Prematuration funded by ISITE Universit\'{e} Lille Nord-Europe.
\end{acknowledgments}


\appendix

\section{\label{Appendix A} Ladder operators}
The far-field asymptotic expressions of the spherical Bessel function and Hankel function of the first kind are, respectively:
\begin{subequations}
\begin{eqnarray}
& & j_{n}(k r) \simeq i^{-(n+1)} e^{i k r} / 2 k r+i^{n+1} e^{-i k r} / 2 k r, 
\label{jn}
\\
& & h_{n}^{(1)}(k r) \simeq i^{-(n+1)} e^{i k r} / k r.
\label{hn}
\end{eqnarray}
\label{asymptotic expressions}
\end{subequations}
and the recursion relation of the spherical Bessel function is
 \begin{equation}
j_{n}'(k r)=(n / k r) j_{n}(k r)-j_{n+1}(k r).
\label{jn recursion}
\end{equation}
where the symbol $'$ means the derivative with respect to $kr$. The ladder operators $L_{\pm}$ has the relationship with the lateral components of the angular momentum operator $L_{x,y}$: $L_{\pm}=L_{x} \pm i L_{y}$ \cite{arfken2013mathematical}. The recursion relations of ladder operators $L_{\pm}$ (or axial component of angular momentum operator $L_z$) and normalized spherical harmonics are \cite{jackson1999classical}
 
 \begin{subequations}
\begin{eqnarray}
L_{+} Y_{n}^{m}=b_n^{-m} Y_{n}^{m+1}, \label{L+}
\\
L_{-} Y_{n}^{m}=b_n^m Y_{n}^{m-1}, \label{L-}
\\
L_{z} Y_{n}^{m}=m Y_{n}^{m}. \label{Lz}
\end{eqnarray}
\label{L Ynm}
\end{subequations}

Finally, the orthogonality relation of the normalized spherical harmonics is \cite{arfken2013mathematical}

\begin{equation}
\int_{0}^{2 \pi} d \varphi \int_{0}^{\pi} \sin \theta d \theta Y_{n}^{m*}(\theta, \varphi) Y_{n'}^{m'}(\theta, \varphi)=\delta_{n n'} \delta_{m m'}.
\label{Y_nm orthogonality}
\end{equation}

\section{\label{Appendix B} Derivation of 3D ART in terms of $H_{nm}$}
\subsection{\label{Appendix B1}  Detailed derivation of $T_x$}

Based on the ART formulas of Eq. (\ref{a_nm & H_nm}), the expression of $x$-component of ART is
\begin{eqnarray}
\begin{aligned}
T_{x} =& -\frac{1}{2 \pi^{2} \rho_{0} k^{3} c_{0}^{2}}  \operatorname{Re}\left\{\sum_{n=0}^{\infty} \sum_{m=-n}^{n} \sum_{n'=0}^{\infty} \sum_{m'=-n'}^{n'} \left(1+A_{n}^{m*}\right) A_{n'}^{m'} \right.
\\
& \left. \times H_{n m}^{*}  H_{n' m'} \iint_{S_{0}} Y_{n}^{m*} L_{x} Y_{n'}^{m'} \sin \theta d \theta d \varphi \right\}  
\end{aligned} \label{Tx in Hnm 1}
\end{eqnarray}

Substitute Eqs. (\ref{L+}) and (\ref{L-}) into (\ref{Tx in Hnm 1}) with the relation $L_x = (L_{+}+L_{-})/2$, 
 \begin{eqnarray}
\begin{aligned}
T_{x} =& -\frac{1}{4 \pi^{2} \rho_{0} k^{3} c_{0}^{2}}  \operatorname{Re}\left\{\sum_{n=0}^{\infty} \sum_{m=-n}^{n} \sum_{n'=0}^{\infty} \sum_{m'=-n'}^{n'} \left(1+A_{n}^{m*}\right) A_{n'}^{m'} \right.
\\
& \left. \times H_{n m}^{*}  H_{n' m'}
\iint_{S_{0}} Y_{n}^{m*} (L_{+}+L_{-}) Y_{n'}^{m'} \sin \theta d \theta d \varphi \right\} 
\\
=& -\frac{1}{4 \pi^{2} \rho_{0} k^{3} c_{0}^{2}}   \operatorname{Re}\left\{\sum_{n=0}^{\infty} \sum_{m=-n}^{n} \sum_{n'=0}^{\infty} \sum_{m'=-n'}^{n'-1} \left(1+A_{n}^{m*}\right) A_{n'}^{m'} \right.
\\
& \left. \times H_{n m}^{*}  H_{n' m'}
\iint_{S_{0}} Y_{n}^{m*} {b}_{n'}^{-m'} Y_{n'}^{m'+1} \sin \theta d \theta d \varphi \right. \\
&+ \left. \sum_{n=0}^{\infty} \sum_{m=-n}^{n} \sum_{n'=0}^{\infty} \sum_{m'=-n'+1}^{n'} \left(1+A_{n}^{m*}\right) A_{n'}^{m'}
\right.
\\
& \left. \times H_{n m}^{*}  H_{n' m'} \iint_{S_{0}} Y_{n}^{m*} {b}_{n'}^{m'} Y_{n'}^{m'-1} \sin \theta d \theta d \varphi
\right\}
\end{aligned} \label{Tx in Hnm 2}
\end{eqnarray}
Note that the regimes of $(n',m')$ in the summation symbol is based on the definition of the normalized spherical harmonics (i.e., $Y_{n'}^{m'}$ and $Y_{n'}^{m' \pm 1}$), which are the intersection part and listed in Table \ref{table2}.
\begin{table}[!htbp]
\small
  \caption{Regime of $(n',m')$ in normalized spherical harmonics for derivation of $F_x$ and $F_y$. Note that based on the definition in Eq. (\ref{incident_pressure}), we have $n' \in [0,\infty]$ and $m' \in [-n',n']$.}
  \label{table2}
  \begin{tabular}{c |c | c | c}
\hline
 &  $n'$ &  $m'$ & Intersection \\
\hline 
$Y_{n'}^{m'+1}$ & $n' \in [0,\infty]$ & $ m' \in [-n'-1, n'-1]$ & $n' \in [0,\infty]$, $ m' \in [-n', n'-1]$\\

$Y_{n'}^{m'-1}$ & $n' \in [0,\infty]$ & $ m' \in [-n'+1, n'+1]$ & $n' \in [0,\infty]$, $ m' \in [-n'+1, n']$\\

 \hline
  \end{tabular}
\end{table}

Using the orthogonality relationship in Eq. (\ref{Y_nm orthogonality}), the expression of $T_x$ is
\begin{eqnarray}
\begin{aligned}
T_{x} =& -\frac{1}{4 \pi^{2} \rho_{0} k^{3} c_{0}^{2}}   \operatorname{Re}\left\{\sum_{n=0}^{\infty} \sum_{m=-n}^{n} \sum_{n'=0}^{\infty} \sum_{m'=-n'}^{n'-1} \left(1+A_{n}^{m*}\right) A_{n'}^{m'}  H_{n m}^{*}  H_{n' m'} b_{n'}^{-m'} \delta_{nn'} \delta_{m,m'+1}  \right. \\
&+ \left. \sum_{n=0}^{\infty} \sum_{m=-n}^{n} \sum_{n'=0}^{\infty} \sum_{m'=-n'+1}^{n'} \left(1+A_{n}^{m*}\right) A_{n'}^{m'}  H_{n m}^{*}  H_{n' m'} b_{n'}^{m'} \delta_{nn'} \delta_{m,m'-1} \right\}
\\
=& -\frac{1}{4 \pi^{2} \rho_{0} k^{3} c_{0}^{2}}    \operatorname{Re}\left\{  \sum_{n=0}^{\infty} \sum_{m=-n+1}^{n} \left(1+A_{n}^{m*}\right) A_{n}^{m-1}  H_{n m}^{*}  H_{n, m-1}  b_{n}^{-m+1} \right. \\
&+ \left. \sum_{n=0}^{\infty} \sum_{m=-n}^{n-1} \left(1+A_{n}^{m*}\right) A_{n}^{m+1}  H_{n m}^{*}  H_{n, m+1} b_{n}^{m+1} \right\}
\end{aligned} \label{Tx in Hnm 3}
\end{eqnarray}
Here, we use a re-index for the second part of Eq. (\ref{Tx in Hnm 3}) by using a variable substitution $q=m+1 \in [-n+1,n]$, and note that $b_{n}^{-m+1} = b_{n}^{m}$, 

\begin{eqnarray}
\begin{aligned}
T_{x} =& -\frac{1}{4 \pi^{2} \rho_{0} k^{3} c_{0}^{2}}    \operatorname{Re}\left\{  \sum_{n=0}^{\infty} \sum_{m=-n+1}^{n} \left(1+A_{n}^{m*}\right) A_{n}^{m-1}  H_{n m}^{*}  H_{n, m-1}  b_{n}^{m} \right. \\
&+ \left. \sum_{n=0}^{\infty} \sum_{q=-n+1}^{n} \left(1+A_{n}^{q-1*}\right) A_{n}^{q}  H_{n, q-1}^{*}  H_{n, q} b_{n}^{q} \right\}
\\
=& -\frac{1}{4 \pi^{2} \rho_{0} k^{3} c_{0}^{2}}    \operatorname{Re}\left\{  \sum_{n=0}^{\infty} \sum_{m=-n+1}^{n} \left(1+A_{n}^{m*}\right) A_{n}^{m-1}  H_{n m}^{*}  H_{n, m-1}  b_{n}^{m} \right. \\
&+ \left. \sum_{n=0}^{\infty} \sum_{m=-n+1}^{n} \left(1+A_{n}^{m-1*}\right) A_{n}^{m}  H_{n, m-1}^{*}  H_{n, m} b_{n}^{m} \right\}
\\
=& -\frac{1}{4 \pi^{2} \rho_{0} k^{3} c_{0}^{2}}    \operatorname{Re}\left\{  \sum_{n=0}^{\infty} \sum_{m=-n+1}^{n} b_{n}^{m} \left(A_{n}^{m-1}+2 A_{n}^{m-1} A_{n}^{m*} + A_{n}^{m*} \right)  H_{n m}^{*}  H_{n, m-1} 
\right\}
\end{aligned} \label{Tx in  Hnm 4}
\end{eqnarray}
which is Eq. (\ref{Tx}) in Sec. \ref{sec: Theoretical method}. Note that Re$\{X\}$ $=$ Re$\{X^*\}$ with $X$  an arbitrary complex number.
\subsection{\label{Appendix B2}  Derivation of $T_y$}

The expression of $y$-component of ART is
\begin{eqnarray}
\begin{aligned}
T_{y} =& -\frac{1}{2 \pi^{2} \rho_{0} k^{3} c_{0}^{2}}  \operatorname{Re}\left\{\sum_{n=0}^{\infty} \sum_{m=-n}^{n} \sum_{n'=0}^{\infty} \sum_{m'=-n'}^{n'} \left(1+A_{n}^{m*}\right) A_{n'}^{m'} \right.
\\
& \left. \times H_{n m}^{*}  H_{n' m'} \iint_{S_{0}} Y_{n}^{m*} L_{y} Y_{n'}^{m'} \sin \theta d \theta d \varphi \right\}  
\end{aligned} \label{Ty in Hnm 1}
\end{eqnarray}

As similar as the derivation for $T_x$, the final expression of $T_y$ in terms of $H_{nm}$ can be obtained by using Eqs. (\ref{L+}) and (\ref{L-}) into (\ref{Ty in Hnm 1}) and $L_y = (L_{+}-L_{-})/{2i}$ instead of $L_x$, as given in Eq. (\ref{Ty}) and omitted here for brevity.

\subsection{\label{Appendix B3}  Detailed derivation of $T_z$}

The expression of $z$-component of ART is
\begin{eqnarray}
\begin{aligned}
T_{z} =& -\frac{1}{2 \pi^{2} \rho_{0} k^{3} c_{0}^{2}}  \operatorname{Re}\left\{\sum_{n=0}^{\infty} \sum_{m=-n}^{n} \sum_{n'=0}^{\infty} \sum_{m'=-n'}^{n'} \left(1+A_{n}^{m*}\right) A_{n'}^{m'} \right.
\\
& \left. \times H_{n m}^{*}  H_{n' m'} \iint_{S_{0}} Y_{n}^{m*} L_{z} Y_{n'}^{m'} \sin \theta d \theta d \varphi \right\}  
\end{aligned} \label{Tz in Hnm 1}
\end{eqnarray}

Inserting Eq. (\ref{Lz}) into (\ref{Tz in Hnm 1}) and using the orthogonality relation in Eq. (\ref{Y_nm orthogonality}), the final expression of $T_z$ in terms of $H_{nm}$ can be derived as
\begin{eqnarray}
\begin{aligned}
T_{z} =& -\frac{1}{2 \pi^{2} \rho_{0} k^{3} c_{0}^{2}}  \operatorname{Re}\left\{\sum_{n=0}^{\infty} \sum_{m=-n}^{n} \sum_{n'=0}^{\infty} \sum_{m'=-n'}^{n'}  \left(1+A_{n}^{m*}\right) A_{n'}^{m'} \right.
\\
& \left. \times H_{n m}^{*}  H_{n' m'} \iint_{S_{0}} Y_{n}^{m*} m' Y_{n'}^{m'} \sin \theta d \theta d \varphi \right\}
\\
=& -\frac{1}{2 \pi^{2} \rho_{0} k^{3} c_{0}^{2}}  \operatorname{Re}\left\{\sum_{n=0}^{\infty} \sum_{m=-n}^{n} \sum_{n'=0}^{\infty} \sum_{m'=-n'}^{n'}  \left(1+A_{n}^{m*}\right) A_{n'}^{m'} H_{n m}^{*}  H_{n' m'} m' \delta_{nn'} \delta_{m m'} \right\}
\\
=& -\frac{1}{2 \pi^{2} \rho_{0} k^{3} c_{0}^{2}}  \operatorname{Re}\left\{\sum_{n=0}^{\infty} \sum_{m=-n}^{n}  
m \left(1+A_{n}^{m*}\right) A_{n}^{m} H_{n m}^{*}  H_{n m}   \right\}
\end{aligned} \label{Tz in Hnm 2}
\end{eqnarray}
which is Eq. (\ref{Tz}) in Sec. \ref{sec: Theoretical method}.

\section{\label{Appendix C} Theoretical derivation of on- and off-axis cylindrical Bessel beam based on ASM}
\subsection{\label{Appendix C1} $H_{nm}$ for an on-axis cylindrical Bessel beam (CBB)}
The $H_{nm}$ coefficients for an on axis particle insonified by a cylindrical Bessel beam were calculated analytically with the angular spectrum method by Sapozhnikov \& Bailey in ref. \citep{sapozhnikov2013radiation}. However, they only provided the result in the paper. Since the calculation is not straightforward, we give here the main elements of the demonstration before extending it to the case of off-axis CBB. The expression of a CBB is given by:

\begin{equation}
p_{i}(x, y, z)=p_{0} e^{i k_{\parallel} z } J_{M}\left(k_{\perp} R \right) e^{i M \varphi}
\label{on axis CBB}
\end{equation}
with $k_{\parallel} = k \cos \beta$, $k_{\perp} = k \sin \beta$, $k = \omega / c_o$ the wavenumber, $\beta$ the cone angle, $R = \sqrt{x^2 + y^2}$ the radius in cylindrical coordinates and $\varphi = \arctan (y/x)$ in $\mathbb{R}^3$ space and $(O,\mathbf{e_z})$ is the central axis of the Bessel beam . In Cartesian coordinates, the angular spectrum in a plane (x,y) of arbitrary altitude z orthogonal to the Bessel beam central axis can be written as:
\begin{equation}
S(k_x,k_y) = \int_{x=- \infty}^{+ \infty} \int_{y = - \infty}^{+\infty} dx dy \,  p_i(x,y,z) e^{-i \mathbf{k}.\mathbf{r}} 
\label{Scar}
\end{equation}
with $\mathbf{k} = k_x \, \mathbf{e_x}  + k_y \, \mathbf{e_y}  + k_z \, \mathbf{e_z}$, $k_z = \sqrt{k^2 - k_x^2 - k_y^2}$, and the expression of the position vector is $\mathbf{r} = x \, \mathbf{e_x}  + y \, \mathbf{e_y}  + z \, \mathbf{e_z}$. This expression gives equation (\ref{S(kx,ky)}) when $z=0$ (a condition which can always be fulfilled with a simple change of frame of reference).

In cylindrical coordinates, the angular spectrum can be recast as:
\begin{equation}
S(k_R,\varphi_k)= \int_{\varphi = 0}^{2 \pi} \int_{R = 0}^{+\infty}  R dR d \varphi p_i(R, \varphi, z) e^{- i \mathbf{k}.\mathbf{r}}
\label{Scyl}
\end{equation}
with $\mathbf{k} = k_R \, \mathbf{e_R} (\varphi_k) +  k_z \, \mathbf{e_z}$, $k_R = \sqrt{{k_x}^2 + {k_y}^2}$, $\varphi_k = \arctan (k_y/k_x)$, and the position vector $\mathbf{r} = R \, \mathbf{e_r}(\varphi) + z \, \mathbf{e_z}$. 

\noindent Inserting the expression of the CBB (\ref{on axis CBB}) in equation (\ref{Scyl}) and considering the angular spectrum in the plane $z=0$ gives:
\begin{equation}
S(k_R,\varphi_k)= p_{0} \int_{\varphi = 0}^{2 \pi} \int_{R = 0}^{+\infty} R d R d \varphi J_{M}\left(k_{\perp} R\right) e^{i [M \varphi - k_{R} R \cos \left(\varphi_{k}-\varphi\right) ] }  
\label{on axis CBB1}
\end{equation}
From the integral definition of a cylindrical Bessel function: $J_{\alpha}(x)=\frac{1}{2 \pi} \int_{-\pi}^{\pi} e^{i(\alpha \tau-x \sin \tau)} d \tau$ and the variable substitution $\pi/2 - \phi = \varphi_k - \varphi$ (so that $d \phi = d \varphi$ and $\cos (\pi/2 - \phi) = \sin \phi$), the integral of the expotential function over $d\varphi$ becomes:
\begin{equation}
\int_{0}^{2 \pi}  d \varphi e^{i [M \varphi - k_{R} R \cos \left(\varphi_{k}-\varphi\right) ] } = 2 \pi i^{-M} J_{M}\left(k_{R} R\right) e^{i M \varphi_{k} }
\label{on axis CBB2}
\end{equation}
Inserting Eq.(\ref{on axis CBB2}) into (\ref{on axis CBB1}) and combining the results with the orthogonality relation of cylindrical Bessel beam $\int_{0}^{\infty} x J_{\alpha}(u x) J_{\alpha}(v x) d x = \delta(u-v) / u$ gives:
\begin{eqnarray}
\begin{aligned}
S(k_R ,\varphi_k )=& p_{0} \int_{0}^{+\infty} R d R J_{M}\left(k_{\perp} R\right) \Big[ 2 \pi i^{-M} J_{M}\left(k_{R} R\right) e^{i M \varphi_{k} } \Big]
\\
=& 2 \pi p_{0} i^{-M} \frac{\delta(k_{\perp}-k_R)}{k_{\perp}} e^{i M \varphi_{k} } 
\end{aligned}
\label{on axis CBB3}
\end{eqnarray}
which is Eq.(62) in Ref.\cite{sapozhnikov2013radiation}. Note that the dirac function is $\neq 0$ only when $k_R = k_{\perp}$. Hence, following the definition of $H_{nm}$ in terms of $S(k_x,k_y)$ (Eq.(\ref{Hnm})), the integral over the disk domain ($k_{x}^{2}+k_{y}^{2} \leq k^{2}$ i.e. $k_R^2 < k^2$) degenerates into the integral over a circle corresponding to $k_R = k_{\perp}$, which in cylindrical coordinates becomes:
\begin{equation}
\begin{aligned}
H_{n m}=& \int_{0}^{2 \pi}  k_{\perp} d \varphi_k  \Big[ 2 \pi p_{0} i^{-M} \frac{1}{k_{\perp}} e^{i M \varphi_{k} }  \Big] Y_{n}^{m*} \left(\theta_{k}, \varphi_{k}\right) e^{-i m \varphi_k}   \\
=& 2 \pi^{3/2} p_{0} i^{-m} \sqrt{\frac{(2 n+1) (n-m) !}{(n+m) !}} P_{n}^m(\cos \beta) \delta_{Mm}
\end{aligned}.
\label{on axis CBB4}
\end{equation}
with ${Y_n^m}^* = \sqrt{\frac{(2 n+1)}{4 \pi} \frac{(n-m) !}{(n+m) !}} P_{n}^m(\cos \theta_k)$ the conjugates of the spherical harmonics and $\int_0^{2 \pi} e^{i (M-m) \varphi_k} d\varphi_k = 2 \pi \delta_{Mm}$. This formula agrees with the expression given in Ref.\cite{sapozhnikov2013radiation}. Note that in Fourier space, we have $d k_{x} d k_{y} = R d k_{R} d \varphi_k$, and $\cos\theta_k = \sqrt{1-k_{R}^{2} / k^{2}} = \sqrt{1-k_{\perp}^{2} / k^{2}}  = \cos \beta$.

\subsection{\label{Appendix C2} $H_{nm}$ for an off-axis cylindrical bessel beam (CBB)}
\begin{figure}
\includegraphics[width=8.6cm]{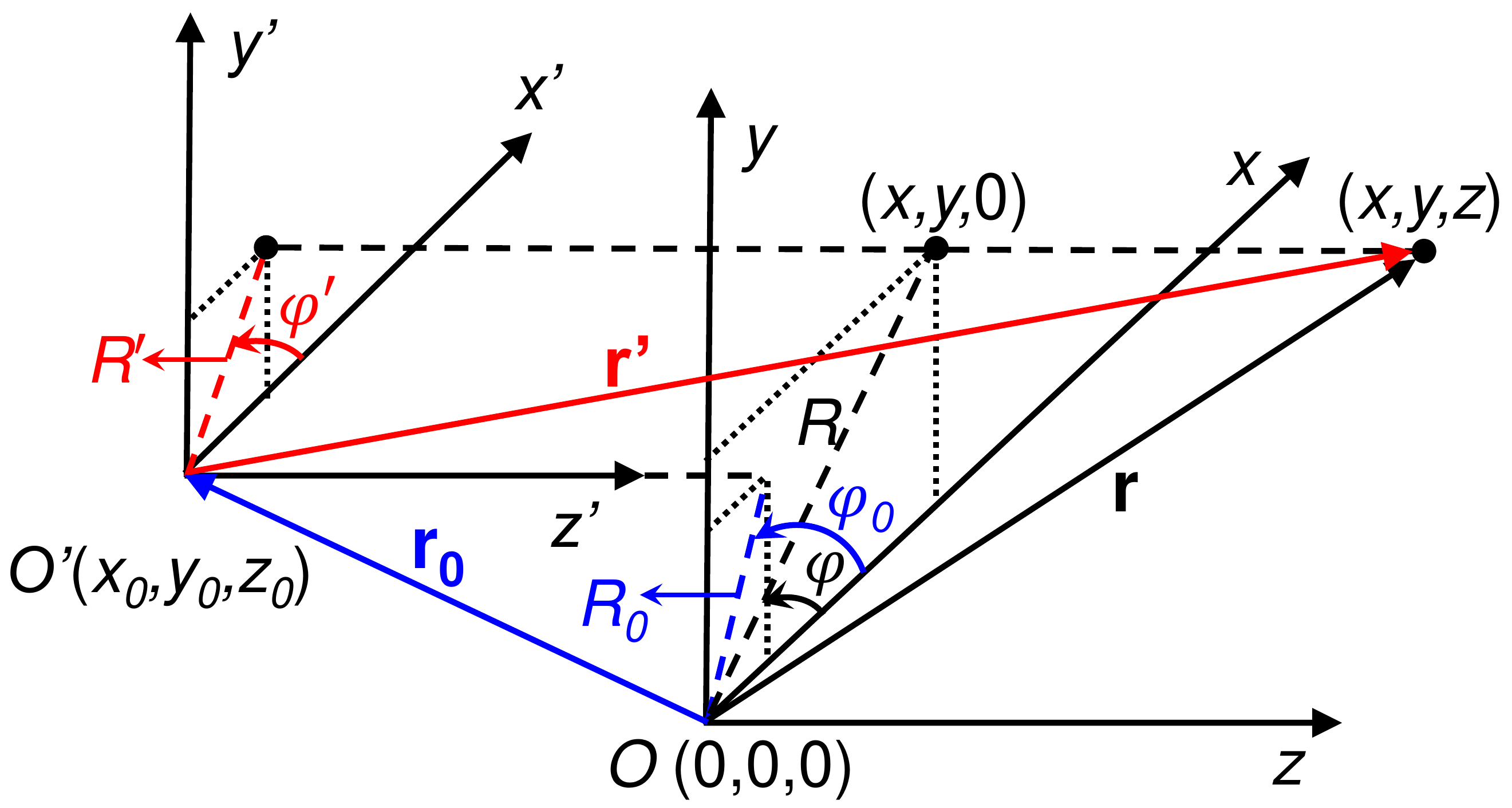}
\caption{\label{Fig2:Coordinates relation} (color online) The global $O(x,y,z)$ with its origin $(0,0,0)$ at the particle center and local coordinates $O'(x',y',z')$ with its origin $(x_0,y_0,z_0)$ at  the beam center.}
\end{figure}

Now let's consider two frames of reference $\mathcal{R}'= (O',(\mathbf{e_x},\mathbf{e_y},\mathbf{e_z}))$, and $\mathcal{R}= (O,(\mathbf{e_x},\mathbf{e_y},\mathbf{e_z}))$, with the point $O'$ located on the central axis of the Bessel beam and the point $O$ corresponding to the "off-axis" center of the particle. Hence the position vectors in the two reference frames are such that:
$\mathbf{r} = \mathbf{r_0} + \mathbf{r'}$, with $\mathbf{r_0} = \mathbf{OO'} = x_0 \, \mathbf{e_x} + y_0 \, \mathbf{e_y} + z_0 \, \mathbf{e_z} = R_o \mathbf{e_r} \, (\varphi_0) + z_0 \, \mathbf{e_z}$. The incident field in $\mathcal{R}'$ is given by:
\begin{equation}
p_{i}(x', y', z')=p_{0} e^{i k_{\parallel} z'} J_{M}\left(k_{\perp} R' \right) e^{i M \varphi'}
\end{equation}
where $(R',\varphi',z')$ are the cylindrical coordinates of $\mathbf{r'}$ in $\mathcal{R}'$.

The angular spectrum $S'$ and $S$ in the planes $(O',(\mathbf{x'},\mathbf{y'})$) and $(O,(\mathbf{e_x},\mathbf{e_y}))$ are respectively given by:
\begin{eqnarray}
& & S'(k_x,k_y) = \int_{x = - \infty}^{+ \infty} \int_{y = - \infty}^{+ \infty} dx' dy'p_i(x',y',z') e^{-i \mathbf{k} . \mathbf{r'}}  \\
& & S(k_x,k_y) = \int_{x = - \infty}^{+ \infty} \int_{y = - \infty}^{+ \infty} dx dy p_i(x-x_0,y-y_0,z-z_0) e^{-i \mathbf{k} . \mathbf{r}} 
\end{eqnarray}
Since (i) the integral over $x$ and $y$ are infinite (and thus not modified by a translation), (ii) $p_i(x,y,z-z_0) = e^{-i k_{\parallel} z_0} p_i(x,y,z)$, (iii) $\mathbf{k}.\mathbf{r} = \mathbf{k}.\mathbf{r'} + \mathbf{k}.\mathbf{r_0} = \mathbf{k}.\mathbf{r'} + k_R \, cos(\varphi_k - \varphi_0)$, we see directly that:
$$
S(k_x,k_y) = S'(k_x,k_y) e^{-i k_{\parallel} z_0 - i k_R cos(\varphi_k - \varphi_0)} 
$$
The on-axis value of the angular spectrum $S'$ was calculated in the previous section (equation (\ref{on axis CBB3})), leading to:

\begin{equation}
S(k_R, \varphi_k) =  2 \pi i^{-M} {p_0}{e^{-{i{k_\parallel }} z_0}} e^{i[ M \varphi_k -  {k_R}{R_0}\cos \left( {{\varphi _k} - {\varphi _0}} \right)]} \frac{\delta(k_\perp - k_R)}{k_\perp}
\label{off axis CBB1}
\end{equation}

We can now proceed very similarly to the on-axis case to compute the $H_{n m}$ coefficients and turn the integral over a disk into an integral over the circle $k_R = k_{\perp}$:
\begin{equation}
\begin{aligned}
H_{n m}=& \int_{0}^{2 \pi}  k_{\perp} d \varphi_k  \Big[2 \pi i^{-M} {p_0}{e^{-{i{k_\parallel }} z_0}} e^{i[ M \varphi_k -  {k_\perp}{R_0}\cos \left( {{\varphi _k} - {\varphi _0}} \right)]} \frac{1}{k_\perp} \Big] Y_{n}^{m*} \left(\theta_{k}, \varphi_{k}\right)
\\
=& 2 \pi i^{-M} p_{0} e^{-i k_\parallel z_{0}} \sqrt{\frac{(2 n+1)}{4 \pi} \frac{(n-m) !}{(n+m) !}} P_{n}^{m}\left(\cos \theta_{k}\right) \int_{\varphi_k=0}^{2 \pi} d \varphi_{k}\left[e^{i\left[(M-m) \varphi_k-k_{\perp} R_{0} \cos \left(\varphi_{k}-\varphi_{0}\right)\right]}\right] 
\end{aligned}.
\label{off axis CBB2}
\end{equation}

Using a variable substitution $\pi/2 - \phi = \varphi_0 - \varphi_k$ (so that $d \varphi_k = d \phi$) to evidence the integral definition of the Bessel function [similar to what was done to obtain Eq. (\ref{on axis CBB2})], Eq. (\ref{off axis CBB2}) turns out to be:
\begin{equation}
H_{n m}=  2 \pi^{3 / 2} i^{-m} p_{0} e^{-i k_{\parallel}z_0} \sqrt{\frac{(2 n+1)(n-m) !}{(n+m) !}} P_{n}^{m}(\cos \beta) e^{i(M-m) \varphi_{0}} J_{m-M}\left(k_{\perp} R_{0}\right)
\label{off axis CBB3}
\end{equation}
with $\cos \theta_k = \cos\beta$, and $J_{M-m}(k_{\perp} R_{0}) = (-1)^{m-M} J_{m-M} (k_{\perp} R_{0}) $ for an integer $(M-m) \in \mathbb{Z}$. When the particle is located on the axis of a CBB [i.e., $(x_0,y_0,z_0) = (0,0,0)$ so that ($R_0, \varphi_{0},z_0) = (0,0,0)]$, Eq. (\ref{off axis CBB3}) degenerates into (\ref{on axis CBB4}). Note that $J_{m-M}(0) = \delta_{mM}$.

For the CBB used in Sec. \ref{sec:validation}, the field is defined by the acoustic potential (with time harmonics omitted) as \cite{gong2017multipole}
\begin{equation}
\Phi_{i}(x, y, z)= \Phi_{0} i^M e^{i k_{\parallel} (z-z_0) } J_{M}\left(k_{\perp} R' \right) e^{i M \varphi'}
\label{off axis CBB4}
\end{equation}
Note that the relation between acoustic potential and pressure is $p = i \omega \rho_0 \Phi$, so that there is a coefficient difference $i^{M+1}$ between Eq. (\ref{H_nm CBB}) and (\ref{off axis CBB3}).

\bibliography{main}

\end{document}